\documentclass[prd,twocolumn,groupedaddress,showpacs,nofootinbib,floatfix]{revtex4}
\usepackage{graphicx}
\usepackage{dcolumn}
\usepackage{amssymb}
\usepackage{mathrsfs}
\usepackage{amsmath}
\usepackage{epsfig}
\usepackage[dvips]{color}
\usepackage{hhline}

\begin{document}

\title{Parametrized leptogenesis from linear seesaw}

\author{Pei-Hong Gu}
\email{peihong.gu@sjtu.edu.cn}

\affiliation{School of Physics and Astronomy, Shanghai Jiao Tong University, 800 Dongchuan Road, Shanghai 200240, China}

\begin{abstract}

We present a purely linear seesaw mechanism in a left-right symmetric framework and then realize a novel leptogenesis scenario for parametrizing the cosmic baryon asymmetty by the charged lepton masses and the light Majorana neutrino mass matrix up to an overall factor. Through the same Yukawa couplings, the lepton-number-conserving decays of the mirror charged leptons can generate three individual lepton asymmetries stored in the ordinary lepton flavors, while the lepton-number-violating processes for the Majorana neutrino mass generation can wash out part of these lepton asymmetries. The remnant lepton asymmetries then can be partially converted to a baryon asymmetry by the sphaleron processes. Our scenario prefers a normal hierarchical neutrino spectrum so that it could be verified by the future data from cosmological observations, neutrino oscillations and neutrinoless double beta decay.

\end{abstract}

\pacs{98.80.Cq, 14.60.Pq, 12.60.Cn, 12.60.Fr}

\maketitle

\section{Introduction}

The precise measurements on the atmospheric, solar, accelerator and reactor neutrinos have established the phenomenon of neutrino oscillations. This fact implies three flavors of neutrinos should be massive and mixed \cite{tanabashi2018}. Meanwhile, the cosmological observations indicate the neutrinos should be extremely light \cite{tanabashi2018}. The tiny but nonzero neutrino masses call for new physics beyond the $SU(3)_c^{}\times SU(2)_L^{}\times U(1)^{}_{Y}$ standard model (SM). Furthermore, the SM is challenged by other puzzles such as the cosmic baryon asymmetry \cite{tanabashi2018}. Currently a seesaw  \cite{minkowski1977,yanagida1979,grs1979,ms1980} extension of the SM has become very attractive since it can simultaneously explain the small neutrino masses and the cosmic baryon asymmetry \cite{fy1986}. In this popular scenario \cite{fy1986,lpy1986,fps1995,ms1998,bcst1999,hambye2001,di2002,gnrrs2003,hs2004,bbp2005}, we do not know much about the masses and couplings involving the non-SM fields. Consequently, we cannot get an exact relation between the cosmic baryon asymmetry and the neutrino mass matrix. For example, we can expect a successful leptogenesis in the canonical seesaw model even if the neutrino mass matrix does not contain any CP phases \cite{di2001}.

In this paper we shall develop a novel leptogenesis \cite{fy1986} scenario where the neutrinos can obtain their tiny masses in the so-called linear \cite{barr2004} seesaw way while the cosmic baryon asymmetry can be parameterized by the neutrino and charged lepton mass matrices up to an overall factor \cite{gh2016}. Our scenario is based on an $SU(3)^{}_{c}\times SU(2)^{}_{L}\times SU(2)^{}_{R}\times U(1)$ left-right symmetric framework\cite{ps1974,mp1975,mp1975-2,ms1975}. Some Yukawa interactions can accommodate the lepton-number-conserving decays of the mirror electron-positron pairs to produce three individual lepton asymmetries in the ordinary lepton flavors although the net lepton number is exactly zero. The same Yukawa interactions can participate in the Majorana neutrino mass generation and then can lead to some lepton-number-violating processes to wash out part of the produced lepton asymmetries. The $SU(2)_L^{}$ sphaleron \cite{krs1985} processes then can partially convert the remnant lepton asymmetries to a baryon asymmetry. Our scenario  prefers a normal hierarchical neutrino spectrum so that it could be verified by the future data from cosmological observations, neutrino oscillations and neutrinoless double beta decay.

\section{The model}

We extend the SM fermions and Higgs scalar, i.e.
\begin{eqnarray}
\!\!\!\!\!\!\!\!\!\!\!\!&&\begin{array}{l}q^{}_{L}(2,1,+\frac{1}{3})(+1),\end{array} \!\!
\begin{array}{l}d^{}_{R}(1,1,-\frac{2}{3})(+1),\end{array}  \!\!
 \begin{array}{l}u^{}_{R}(1,1,+\frac{4}{3})(+1),\end{array} \nonumber\\
[2mm]
\!\!\!\!\!\!\!\!\!\!\!\!&&\begin{array}{l}l^{}_{L}(2,1,-1)(-1),\end{array} \!\!
\begin{array}{l}e^{}_{R}(1,1,-2)(-1);\end{array}\!\!
\begin{array}{l}\phi^{}_L(2,1,-1)(0),\end{array}
\end{eqnarray}
by a mirror copy \cite{bm1989,bm1990,bcs1991,gu2012,cgnr2013,gu2014,gh2016,abbas2016,abbas2017,abbas2017-2,abbas2017-3,gu2017-2},
\begin{eqnarray}
\!\!\!\!\!\!\!\!\!\!\!\!&&\begin{array}{l}Q^{}_{R}(1,2,+\frac{1}{3})(-1),\end{array} \!\!
\begin{array}{l}D^{}_{L}(1,1,-\frac{2}{3})(-1),\end{array} \!\!  
\begin{array}{l}U^{}_{L}(1,1,+\frac{4}{3})(-1),\end{array} \nonumber\\
[2mm]
\!\!\!\!\!\!\!\!\!\!\!\!&&\begin{array}{l}L^{}_{R}(1,2,-1)(+1),\end{array}\!\!
\begin{array}{l}E^{}_{L}(1,1,-2)(+1);\end{array}\!\!
\begin{array}{l}\phi^{}_R(1,2,-1)(0).\end{array}
\end{eqnarray}
Here and thereafter the first brackets following the fields describe the transformations under the $SU(2)^{}_{L}\times SU(2)_R^{}\times U(1)$ gauge groups, while the second brackets denote a $U(1)_{3B-L}^{}$ global symmetry. Our model also contains the following Higgs scalars and fermions,
\begin{eqnarray}
\!\!\!\!\!\!\!\!\!\!\!\!&&\begin{array}{l}\chi^{}_L(2,1,-1)(-2),\end{array}\!\!
\begin{array}{l}\chi^{}_R(1,2,-1)(+2),\end{array}\!\!
\begin{array}{l}\Sigma (2,2,0)(-2),\end{array}\nonumber\\
[2mm]
\!\!\!\!\!\!\!\!\!\!\!\!&&\begin{array}{l}\xi(1,1,0)(+1);\end{array}\!
\!\begin{array}{l}N^{}_L(1,1,0)(-1).\end{array}
\end{eqnarray}
The Higgs bidoublet $\Sigma$ can be expressed in terms of two $SU(2)_L^{}$ doublets, i.e. $\Sigma=\left[\sigma_1^{}~\tilde{\sigma}_2^{}\right]$.

In this work, the discrete left-right symmetry is taken to be the $CP$ (charge conjugation and parity), i.e.
\begin{eqnarray}
&&\phi_L^{}\stackrel{CP}{\longleftrightarrow}\phi_{R}^{\ast}\,, ~~  \chi_L^{}\stackrel{CP}{\longleftrightarrow}\chi_{R}^{\ast}\,,~~\Sigma\stackrel{CP}{\longleftrightarrow}\Sigma^T_{}\,,~~\xi \stackrel{CP}{\longleftrightarrow} \xi\,,\nonumber\\
&&q_{L}^{}\stackrel{CP}{\longleftrightarrow}Q_{R}^{c}\,,~~d_{R}^{}\stackrel{CP}{\longleftrightarrow}D_{L}^{c}\,,~~u_{R}^{}\stackrel{CP}{\longleftrightarrow}U_{L}^{c}\,,\nonumber\\
&&l_{L}^{}\stackrel{CP}{\longleftrightarrow}L_{R}^{c}\,,~~e_{R}^{}\stackrel{CP}{\longleftrightarrow}E_{L}^{c}\,,~~N_{L}^{}\stackrel{CP}{\longleftrightarrow}N_{L}^{}\,.
\end{eqnarray}
The global $U(1)_{3B-L}^{}$ symmetry and the discrete left-right symmetry are both conserved exactly.

For simplicity we do not write down the full scalar potential. Instead we show the cubic terms and some quartic terms as below, 
\begin{eqnarray}
\label{potential}
V&\supset&\kappa \xi^2_{} \left(\phi^\dagger_{L} \chi_L^{} +\phi^T_R \chi^{\ast}_R\right)+\rho \left(\phi^\dagger_L \Sigma \chi^{}_R + \phi^T_R \Sigma^T_{} \chi^{\ast}_L\right) \nonumber\\
&&+\textrm{H.c.}\,.
\end{eqnarray}
As for the Yukawa interactions, they are given by 
\begin{eqnarray}
\label{yukawa}
\!\!\!\!\mathcal{L}_Y^{}\!\!&=&\!\!-y^{}_d \left(\bar{q}^{}_L \tilde{\phi}^{}_{L} d^{}_R + \bar{Q}^c_R \tilde{\phi}^\ast_{R} D^{c}_L  \right)   \nonumber\\
\!\!\!\!\!\!&&\!\!-y^{}_u \left(\bar{q}^{}_L \phi^{}_{L} u^{}_R + \bar{Q}^c_R  \phi^\ast_{R} U^{c}_L  \right)- y_Q^{}\bar{q}^{}_L \tilde{\Sigma} Q_R^{}\nonumber\\
\!\!\!\!\!\!&&\!\!-y^{}_e \left(\bar{l}^{}_L \tilde{\phi}^{}_{L} e^{}_R + \bar{L}^c_R \tilde{\phi}^\ast_{R} E^{c}_L  \right)\nonumber\\
\!\!\!\!\!\!&&\!\!-y^{}_N \left(\bar{l}^{}_L \chi^{}_{L} N^{c}_L + \bar{L}^c_R  \chi^\ast_R N^{c}_L  \right) - y_L^{} \bar{l}_L^{} \Sigma L_R^{}+\textrm{H.c.}\nonumber\\
\!\!\!\!\!\!&&\!\! \textrm{with}~~y_Q^{}=y_Q^T\,,~~y_L^{}=y_L^T\,.
\end{eqnarray}
Note the $U(1)_{3B-L}^{}$ global symmetry has forbidden the gauge-invariant mass terms of the $[SU(2)]$-singlet fermions.

\section{Fermion masses}

From the full potential which are not shown for simplicity, we can expect the VEVs to be
\begin{eqnarray}
\langle\xi\rangle\,, ~ \langle\chi_R^{0}\rangle \,,~\langle\phi_R^{0} \rangle \gg \langle\phi_L^{0}\rangle \,,~\langle\chi_L^{0} \rangle \,, ~  \langle\sigma_{1,2}^{0} \rangle \,.
\end{eqnarray}
The Yukawa interactions (\ref{yukawa}) then can reasonably yield  
\begin{eqnarray}
&&y_Q^{}  \langle\sigma_{2}^{0} \rangle  \ll y_{d}^{} \langle\phi_R^{0} \rangle\,, ~~y_Q^{}  \langle\sigma_{1}^{0} \rangle  \ll y_{u}^{} \langle\phi_R^{0} \rangle\,, \nonumber\\
[2mm]
&&y_L^{}  \langle\sigma_{2}^{0} \rangle  \ll y_{e}^{} \langle\phi_R^{0} \rangle\,,~~y_L^{}  \langle\sigma_{1}^{0} \rangle  \ll y_{N}^{} \langle\chi_R^{0} \rangle\,. 
\end{eqnarray}
This means we can safely ignore the mixing between the ordinary charged fermions $(f=d,u,e)$ and their mirror partners $(F=D,U,E)$. Thus the mass eigenstates of the charged fermions can come from
\begin{eqnarray}
\mathcal{L}&\supset&  - m_f^{} \bar{f}_L^{} f_R^{} - M_F^{}  \bar{F}_R^{c} F_L^{c}+ \textrm{H.c.}~~\textrm{with}\nonumber\\
[2mm]
&&m_f^{}=y_{f}^{} \langle\phi_L^{0} \rangle \,,~~M_F^{}=y_{f}^{} \langle\phi_R^{0} \rangle\,.
\end{eqnarray} 
Meanwhile, we can apply the linear seesaw mechanism to the neutral fermions, i.e. 
\begin{eqnarray}
\label{linear}
\mathcal{L}&\supset& \! -\left[\bar{\nu}_L^{}~ \bar{N}_L^{} ~ \bar{N}_R^{c}\right] \!\!\left[\begin{array}{ccc} 0 &y_N^{} \langle \chi_L^0 \rangle & y_L^{} \langle \sigma_1^0 \rangle \\
[1mm]
y_N^T \langle \chi_L^0\rangle & 0& y_N^{T} \langle \chi_R^0 \rangle \\
[1mm]
Y_L^T\langle \sigma_1^0\rangle & y_N^{} \langle \chi_R^0 \rangle  & 0  \end{array}\right]\!\!\! \left[\begin{array}{c}\nu_L^{c}\\
[1mm]
N_L^{c}\\
[1mm]
N_R^{}\end{array}\right]\nonumber\\
&&\!+\textrm{H.c.}\nonumber\\
&\simeq &\!- M_N^{} \bar{N}_R^{c}N_L^{c} -\frac{1}{2} m_\nu^{} \bar{\nu}_L^{} \nu_L^c +\textrm{H.c.}~~\textrm{with}\nonumber\\
&&\!M_N^{}=y_{N}^{} \langle\chi_R^{0} \rangle\,,~~m_\nu^{} =m_\nu^T= -2 y_L^{}  \langle\sigma_1^{0} \rangle \frac{ \langle\chi_L^{0} \rangle}{ \langle\chi_R^{0} \rangle}\,.
\end{eqnarray}

Note the VEVs $\langle\phi_{L}^{0} \rangle$, $\langle\chi_{L}^{0} \rangle$ and $\langle\sigma_{1,2}^{0} \rangle$,  should be constrained by 
\begin{eqnarray}
 &&\sqrt{ \langle\phi_{L}^{0} \rangle^2_{} + \langle\chi_{L}^{0} \rangle^2_{} +  \langle\sigma_{1}^{0} \rangle^2_{} +  \langle\sigma_{2}^{0} \rangle^2_{}}\equiv v =174\,\textrm{GeV}\,,\nonumber\\
 [2mm]
&& \langle\phi_{L}^{0} \rangle = \frac{m_t^{}}{y_t^{}}> \frac{m_t^{}}{\sqrt{4\pi}}\simeq 48.5\,\textrm{GeV}\,,
 \end{eqnarray} 
which implies 
\begin{eqnarray}
 \mu_2^{2}&\equiv&2\langle\chi_{L}^{0} \rangle \langle\sigma_{1}^{0}\rangle \leq \langle\chi_{L}^{0} \rangle ^2_{}+  \langle\sigma_{1}^{0} \rangle^2_{} < \left(174^2- 48.5^2\right) \,\textrm{GeV}^2_{} \nonumber\\
 &=&2.79\times 10^4_{}\,\textrm{GeV}^2_{}\,.
  \end{eqnarray} 
In addition, the tiny but nonzero neutrino masses require
\begin{eqnarray}
\mu_1^{}\equiv \frac{ 2\langle\chi_{L}^{0} \rangle \langle\sigma_{1}^{0}\rangle}{ \langle\chi_{R}^{0} \rangle} &=& \frac{ \hat{m}_\nu^{}}{\hat{y}_L^{}}>  \frac{ m_{\textrm{max}}^{}}{\sqrt{4\pi}}=0.014\,\textrm{eV}\left(\frac{m_{\textrm{max}}^{}}{0.05\,\textrm{eV}}\right)\,,\nonumber\\
&&
\end{eqnarray} 
with $ m_{\textrm{max}}^{}$ being the largest eigenvalue of the neutrino mass matrix,
\begin{eqnarray}
m_\nu^{}=U\hat{m}_\nu^{} U^T_{} = U \textrm{diag}\{m_1^{}, m_2^{}, m_3^{}\} U^T_{}\,.
\end{eqnarray} 
Here the PMNS matrix $U$ contains three mixing angles, one Dirac phase and two Majorana phases, i.e.
\begin{eqnarray}
\label{pmns}
\!\!\!\!\!\!U\!\!\!&=&\!\!\!\left[\!\!\begin{array}{ccl}
\!\!c_{12}^{}c_{13}^{}& \!\!s_{12}^{}c_{13}^{}&\!\!  s_{13}^{}e^{-i\delta}_{}\\
[2mm] \!\!-\!s_{12}^{}c_{23}^{}\!-\!c_{12}^{}s_{23}^{}s_{13}^{}e^{i\delta}_{}
&\!\!~~c_{12}^{}c_{23}^{}\!-\!s_{12}^{}s_{23}^{}s_{13}^{}e^{i\delta}_{}
&\!\! s_{23}^{}c_{13}^{}\\
[2mm]\!\! ~~s_{12}^{}s_{23}^{}\!-\!c_{12}^{}c_{23}^{}s_{13}^{}e^{i\delta}_{}
& \!\!-\!c_{12}^{}s_{23}^{}\!-\!s_{12}^{}c_{23}^{}s_{13}^{}e^{i\delta}_{}
& \!\!c_{23}^{}c_{13}^{}
\end{array}\!\!\right]\!\!.\nonumber\\
\!\!\!\!\!\!&&\!\!\!\times \textrm{diag}\{e^{i\alpha_1^{}/2}_{}, e^{i\alpha_2^{}/2}_{}, 1\}\,.
\end{eqnarray}

\begin{figure*}
\vspace{6.5cm} \epsfig{file=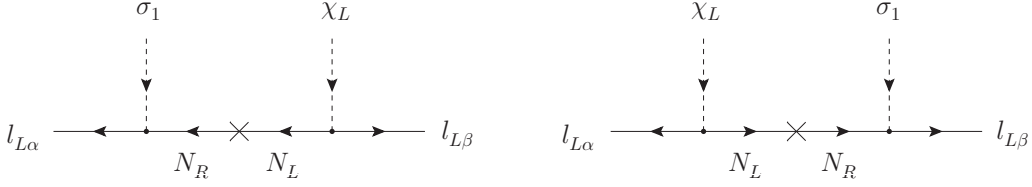, bbllx=7.5cm, bblly=6.0cm,
bburx=17.5cm, bbury=16cm, width=7cm, height=7cm, angle=0,
clip=0} \vspace{-10.5cm} \caption{\label{numass} The lepton-number-violating processes for generating the Majorana neutrino masses.}
\end{figure*}

The linear seesaw can be also understood at the electroweak level. From the Yukawa interactions (\ref{yukawa}), we can read
\begin{eqnarray}
\label{yukawaew}
\mathcal{L}\!\!&\supset&\!\!   - \hat{y}_e^{}\bar{l}_L^{}\tilde{\phi}_L^{} e_R^{}   - \hat{M}_E^{}  \bar{E}_R^{c} E_L^{c}  - X_R^{} \hat{y}_N^{}   \bar{l}_L^{}\chi_L^{} N_L^{c} \nonumber\\
\!\!&&\!\!
-\hat{M}_N^{} \bar{N}_R^c N_L^c
-y_L^{} X_R^{\ast}  \bar{l}_L^{}\sigma_1^{} N_R^{}-y_L^{}\bar{l}_L^{}\tilde{\sigma}_2^{} E_R^{}+\textrm{H.c.}\,\nonumber\\
\!\!&&\!\!\textrm{with}~y_N^{} = X_R^{} \hat{y}_N^{} X_L^\dagger\,,~X_{L,R}^\dagger X_{L,R}^{}=X_{L,R}^{} X_{L,R}^\dagger =1\,.\nonumber\\
\!\!&&\!\!
\end{eqnarray}
Here the yukawa couplings $y_{e}^{}$ and the mass matrices $M_{E,N}^{}$ have been chosen to be diagonal and real without loss of generality. As shown in Fig. \ref{numass}, the left-handed neutrinos $\nu_L^{}$ can acquire their Majorana masses (\ref{linear}) by integrating out the heavy Dirac pairs $N=N_L^{}+N_R^{}$, i.e.
\begin{eqnarray}
\mathcal{L}\!\!&\supset&\!\!  -y_L^{} X_R^{\ast} \frac{1}{\hat{M}_{N}^{}} \hat{y}_N^{} X_R^T   \bar{l}_L^{}\sigma_{1}^{} \chi_L^{T} l_L^c  \nonumber\\
\!\!&&\!\! - X_R^{} \hat{y}_N^{} \frac{1}{\hat{M}_{N}^{}} X_R^\dagger y_L^T   \bar{l}_L^{}\chi_L^{}\sigma_{1}^{T} l_L^c +\textrm{H.c.}\nonumber\\
\!\!&=&\!\! - \frac{y_L^{}}{\langle\chi_R^0\rangle} \left(  \bar{l}_L^{}\sigma_{1}^{} \chi_L^{T} l_L^c + \bar{l}_L^{}\chi_L^{}\sigma_{1}^{T} l_L^c \right)+\textrm{H.c.}\,.
\end{eqnarray}
Note the conditions $y_{N,L}^{}< \sqrt{4\pi}$ and $\mu_2^{2}<2.79\times 10^{4}_{}\,\textrm{GeV}^2_{}$ should constrain the heavy Dirac fermion masses $M_N^{}<7\times 10^{15}_{}\,\textrm{GeV}$ for the light Majorana neutrino mass  $m_{\textrm{max}}^{}\geq 0.05\,\textrm{eV}$.

\section{Lepton and baryon asymmetries}

\begin{figure*}
\vspace{6.5cm} \epsfig{file=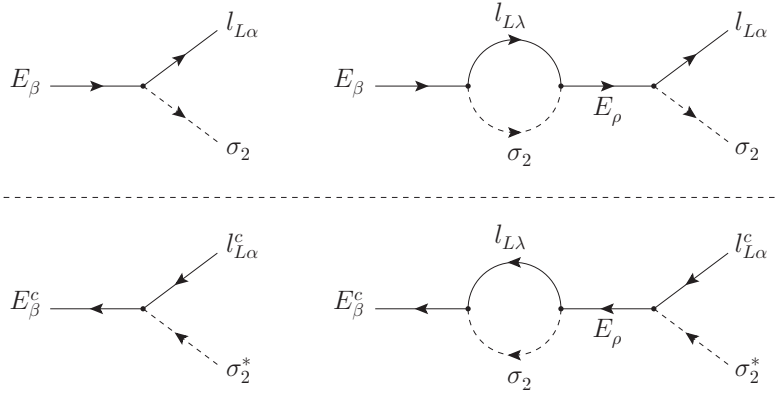, bbllx=6cm, bblly=6.0cm,
bburx=16cm, bbury=16cm, width=7cm, height=7cm, angle=0,
clip=0} \vspace{-7.5cm} \caption{\label{edecay} The lepton-number-conserving decays of the mirror charged leptons into the ordinary leptons.}
\end{figure*}

As shown in Fig. \ref{edecay}, the mirror charged leptons $E_\beta^{}$ can decay into the ordinary lepton doublets $l_{L\alpha}^{}$ and the Higgs doublet $\sigma_2^{}$. These decays can generate three individual lepton asymmetries $L_{e,\mu,\tau}^{}$ stored in the ordinary lepton flavors $l_{Le,L\mu,L\tau}^{}$ if the CP is not conserved, i.e.
\begin{eqnarray}
L_{\beta\alpha}^{}&\propto&\Gamma(E_\beta^{}\rightarrow L_{L\alpha}^{}\sigma_2^{})-\Gamma(E_\beta^{c}\rightarrow L_{L\alpha}^{c}\sigma_2^{\ast})\neq 0\,,\nonumber\\
[2mm]
L_\alpha^{}&=& \sum_{\beta}^{}L_{\beta\alpha}^{}\,.
\end{eqnarray}
We calculate the decay width at tree level, 
\begin{eqnarray}
\Gamma_\beta^{}&\equiv& \Gamma(E_\beta^{}\rightarrow l_{L\alpha}^{} +\sigma_2^{})=\Gamma(E_\beta^{c}\rightarrow l_{L\alpha}^{c} +\sigma_2^{\ast})\nonumber\\
&=&\frac{1}{16\pi} \left(y_L^\dagger y_L^{}\right)_{\beta\beta}^{} M_{E_\beta}^{}\nonumber\\
&=& \frac{1}{16\pi} \frac{\left(m_\nu^\dagger m_\nu^{}\right)_{\beta\beta}^{}}{\mu_1^2} M_{E_\beta}^{}\,,
\end{eqnarray}
and then the CP asymmetry at one-loop level,
\begin{eqnarray}
\varepsilon _{\beta\alpha}^{}&=&\frac{\Gamma(E_\beta^{}\rightarrow l_{L\alpha}^{} +\sigma_2^{})-\Gamma(E_\beta^{c}\rightarrow l_{L\alpha}^{c} +\sigma_2^{\ast})}{\Gamma_\beta^{} }\nonumber\\
&=&\frac{1}{4\pi}\sum_{\rho}^{}\frac{\textrm{Im}\left[\left(y_L^\dagger y_L^{}\right)_{\rho\beta}^{} y_{L\alpha\beta}^{\ast}  y_{L\alpha\rho}^{} \right]}{\left(y_L^\dagger y_L^{}\right)_{\beta\beta}^{}}\frac{M_{E_\beta}^2 }{M_{E_\beta}^2-M_{E_\rho}^2 }\nonumber\\
&=&\frac{1}{4\pi}\sum_{\rho}^{}\frac{\textrm{Im}\left[\left(m_\nu^\dagger m_\nu^{}\right)_{\rho\beta}^{} m_{\alpha\beta}^{\ast}  m_{\alpha\rho}^{} \right]}{ \mu_1^2  \left(m_\nu^\dagger m_\nu^{}\right)_{\beta\beta}^{}}\frac{m_{\beta}^2 }{m_{\beta}^2-m_{\rho}^2 }\,.
\end{eqnarray}
It is easy to check 
\begin{eqnarray}
\varepsilon _{\beta e}^{}+ \varepsilon _{\beta \mu }^{}+ \varepsilon _{\beta\tau}^{}\equiv 0\,,
\end{eqnarray}
as a result of the lepton number conservation.

The decays of the mirror electron-positron pairs should dominate the individual lepton asymmetries $L_{e,\mu,\tau}^{}$ since the mirror electron is much lighter than the mirror muon and tau. When the mirror electrons and positrons go out of equilibrium at a temperature $T_D^{}$, the individual lepton asymmetries $L_{e,\mu,\tau}^{}$ can be  produced, i.e.
\begin{eqnarray}
L_\alpha^{} \simeq \left.\varepsilon _{e\alpha}^{}\left(\frac{n_{E_e}^{\textrm{eq}}}{s}\right)\right|_{T=T_D^{}}^{}\,.
\end{eqnarray}
Here $n_{E_e}^{\textrm{eq}}$ is the equilibrium number density and $s$ is the entropy density. For the following demonstration, we specify the decay width,
\begin{eqnarray}
\Gamma_{e}^{}
&=& \frac{1}{16\pi} \frac{\left(m_\nu^\dagger m_\nu^{}\right)_{ee}^{}}{\mu_1^2} M_{E_e}^{}\,,
\end{eqnarray}
and the CP asymmetries,
\begin{eqnarray}
\varepsilon _{ee}^{}
&\simeq&-\frac{1}{4\pi}\frac{\textrm{Im}\left[\left(m_\nu^\dagger m_\nu^{}\right)_{\mu e}^{} m_{e e}^{\ast}  m_{e \mu}^{} \right]}{ \mu_1^2  \left(m_\nu^\dagger m_\nu^{}\right)_{ee}^{}}\frac{m_{e}^2 }{m_{\mu}^2}\,,\nonumber\\
\varepsilon _{e\mu}^{}
&\simeq&-\frac{1}{4\pi}\frac{\textrm{Im}\left[\left(m_\nu^\dagger m_\nu^{}\right)_{\mu e}^{} m_{\mu e}^{\ast}  m_{\mu \mu}^{} \right]}{ \mu_1^2  \left(m_\nu^\dagger m_\nu^{}\right)_{ee}^{}}\frac{m_{e}^2 }{m_{\mu}^2}\,,\nonumber\\
\varepsilon _{e\tau}^{}
&\simeq&-\frac{1}{4\pi}\frac{\textrm{Im}\left[\left(m_\nu^\dagger m_\nu^{}\right)_{\mu e}^{} m_{\tau e}^{\ast}  m_{\tau \mu}^{} \right]}{ \mu_1^2  \left(m_\nu^\dagger m_\nu^{}\right)_{ee}^{}}\frac{m_{e}^2 }{m_{\mu}^2}\,.
\end{eqnarray}

On the other hand, the model provides the lepton-number-violating interactions for generating the Majorana neutrino masses, as shown in Fig. \ref{numass}. The interaction rates of the related lepton-number-violating processes are computed by \cite{fy1990}
\begin{eqnarray}
\Gamma_{\alpha\beta^{}}^{}=\frac{1}{\pi^3_{}} \frac{\left| m_{\alpha\beta}^{}\right|^2_{} T^3_{}}{v^4_{}}~~\textrm{for}~~T<M_N^{}\,.
\end{eqnarray}
At the temperature,
\begin{eqnarray}
\label{tlnv}
\!\!\!\!\!\!\!\!T_{\alpha\beta}^{}&=&\left(\frac{8\pi^{9}_{}g_{\ast}^{}}{90}\right)^{\!\frac{1}{2}}_{}\frac{v^4_{}}{M_{\textrm{Pl}}^{}\left| m_{\alpha\beta}^{}\right|^2_{} } \nonumber\\
\!\!\!\!\!\!\!\!&=&1.63\times10^{13}_{}\,\textrm{GeV}\left(\frac{0.05\,\textrm{eV}}{\left|m_{\alpha\beta}^{}\right|}\right)^2_{}\,,
\end{eqnarray}
these lepton-number-violating processes will begin to decouple, i.e. 
\begin{eqnarray}
\left.\left[\Gamma_{\alpha\beta^{}}^{}< H(T)=\left(\frac{8\pi^{3}_{}g_{\ast}^{}}{90}\right)^{\frac{1}{2}}_{}\frac{T^2_{}}{M_{\textrm{Pl}}^{}}\right]\right|_{T<T_{\alpha\beta}^{}}^{}\,.
\end{eqnarray}
Here $H(T)$ is the Hubble constant with $M_{\textrm{Pl}}^{}\simeq 1.22\times 10^{19}_{}\,\textrm{GeV}$ being the Planck mass and $g_{\ast}^{}=110.75$ being the relativistic degrees of freedom (the SM fields plus one additional Higgs doublet $\sigma_{2}^{}$).

We thus can expect only the lepton asymmetry stored in certain lepton flavor(s) can survive from the lepton-number-violating processes, i.e.
\begin{eqnarray}
L=\left\{\begin{array}{lll}L_e^{}&\textrm{for}&T_{\mu\mu,\mu\tau,\tau\tau}^{}<T_D^{}<T_{ee,e\mu,e\tau}^{}\,,\\
[1mm]
L_\mu^{}&\textrm{for}&T_{ee,e\tau,\tau\tau}^{}<T_D^{}<T_{e\mu,\mu\mu,\mu\tau}^{}\,,\\
[1mm]
L_\tau^{}&\textrm{for}&T_{ee,e\mu,\mu\mu}^{}<T_D^{}<T_{e\tau,\mu\tau,\tau\tau}^{}\,,\\
[1mm]
L_e^{}+L_\mu^{}&\textrm{for}&T_{\tau\tau}^{}<T_D^{}<T_{ee,e\mu,e\tau,\mu\mu,\mu\tau}^{}\,,\\
[1mm]
L_e^{}+L_\tau^{}&\textrm{for}&T_{\mu\mu}^{}<T_D^{}<T_{ee,e\mu,e\tau,\mu\tau,\tau\tau}^{}\,,\\
[1mm]
L_\mu^{}+L_\tau^{}&\textrm{for}&T_{ee}^{}<T_D^{}<T_{e\mu,e\tau,\mu\mu,\mu\tau,\tau\tau}^{}\,.\end{array}\right.
\end{eqnarray}
The $SU(2)_L^{}$ sphaleron processes then can partially transfer the remnant lepton asymmetry $L$ to a baryon asymmetry $B$. From Eq. (\ref{tlnv}), the lepton-number-violating processes can go out of equilibrium before the sphalerons become active, i.e.
\begin{eqnarray}
T_{\alpha\beta}^{}> T_{\textrm{sph}}^{}\simeq 10^{12}_{}\,\textrm{GeV}\,.
\end{eqnarray}
The final baryon asymmetry $B$ then can be given by \cite{ht1990}
\begin{eqnarray}
B= -\frac{28}{79} L\,.
\end{eqnarray}

\section{Numerical example}

In the weak washout region \cite{kt1990}, i,e,
\begin{eqnarray}
 K=\frac{\Gamma_e^{}}{2H}\left|_{T=M_{E_e}^{}}^{}\right.< 1\,,
\end{eqnarray}
we can roughly estimate \cite{kt1990}
\begin{eqnarray}
T_D^{} \sim M_{E_e}^{}\sqrt{K}\,,~~L_\alpha^{} \sim \frac{\varepsilon_{e\alpha}}{g_\ast^{}}\,.
\end{eqnarray}
In this case, as shown in Eqs. (21), (22), (24), (26), (28) and (30), the final baryon asymmetry can be fully determined by the charged lepton masses and the Majorana neutrino mass matrix, up to an overall factor depending on the  parameters $\mu_{1,2}^{}$ and $M_{E_e}^{}$.

We have known the charged lepton masses $m_e^{}=511\,\textrm{keV}$, $m_\mu^{}=107\,\textrm{MeV}$, and the normal(inverted) neutrino parameters $\Delta m_{21}^2=m_2^2-m_1^2=7.37\times 10^{-5}_{}\,\textrm{eV}^2_{}$, $\Delta m_{31}^2(\Delta m_{23}^2)=m_3^2- m_1^2(m_2^2-m_3^2)=2.56(2.54)\times 10^{-3}_{}\,\textrm{eV}^2_{}$, $\sin ^2_{}\theta_{12}^{}=0.297$, $\sin ^2_{}\theta_{23}^{}=0.425(0.589)$, $\sin ^2_{}\theta_{13}^{}=0.0215(0.0216)$ \cite{tanabashi2018}. It seems difficult for the inverted hierarchical and quasi-degenerate neutrinos to fulfil the conditions in Eq. (28). So, we consider the normal hierarchical neutrinos to give a numerical example. Specifically, we fix $m_1^{}=0$, $\alpha_2^{}=\pi/2$, $\delta=3\pi/2$, and then take $\mu_1^{}=0.057\,\textrm{eV}$, $\mu_2^2= 2.79\times 10^{4}_{}\,\textrm{GeV}^2_{}$, $M_{E_e}^{}=6.8\times 10^{14}_{}\,\textrm{GeV}$. With these inputs, we obtain $K\simeq 0.96$, $\varepsilon_{ee}^{}\simeq -3.3\times 10^{-8}_{} $, $T_D^{}\sim 6.7\times 10^{14}_{}\,\textrm{GeV}$, $T_{\tau\tau}^{}\simeq 5.0\times 10^{13}_{}\,\textrm{GeV}$, $T_{\mu\tau}^{}\simeq 6.8\times 10^{13}_{}\,\textrm{GeV}$, $T_{\mu\mu}^{}\simeq 9.3\times 10^{13}_{}\,\textrm{GeV}$, $T_{e\mu}^{}\simeq 6.8\times 10^{14}_{}\,\textrm{GeV}$, $T_{e\tau}^{}\simeq 4.4\times 10^{15}_{}\,\textrm{GeV}$, $T_{ee}^{}\simeq 5.5\times 10^{15}_{}\,\textrm{GeV}$ and hence $B \sim 10^{-10}_{}$.

 \section{Conclusion} In this paper we have demonstrated a novel linear seesaw scenario for paramerizing the cosmic baryon asymmetty by the charged lepton masses and the light Majorana neutrino mass matrix up to an overall factor. Through the lepton-number-conserving decays of the mirror electron-positron pairs, we can obtain three individual lepton asymmetries stored in the ordinary lepton flavors although the total lepton asymmetrt is exactly zero. The lepton-number-violating processes for the neutrino mass generation can wash out the lepton asymmetry stored in certain ordinary lepton flavor(s). Remarkably, these lepton-number-conserving and lepton-number-violating interactions originate from the same Yukawa couplings. The remnant lepton asymmetry can be partially converted to a baryon asymmetry by the sphaleron processes. Our scenario seems difficult to work for an inverted hierarchical or a quasi-degenerate neutrino spectrum. Instead, it prefers to the normal hierarchical neutrinos. This means our scenario can be ruled out if the future cosmological observations, neutrino oscillations and neutrinoless double beta decay  confirm the inverted hierarchical or quasi-degenerate neutrino spectrum.

 \textbf{Acknowledgement}: This work was supported in part by the NSFC (Grant No. 11675100) and also in part by the Recruitment Program for Young Professionals (Grant No. 15Z127060004).


\begin{thebibliography}{99}




\bibitem{tanabashi2018}
M. Tanabashi {\it et al.}, (Particle Data Group), Phys. Rev. D \textbf{98}, 030001 (2018).




\bibitem{minkowski1977}
P. Minkowski, Phys. Lett. B \textbf{67}, 421 (1977).

\bibitem{yanagida1979}
T. Yanagida, {\it Proceedings of the Workshop on Unified Theory and the Baryon Number of the Universe}, ed. O. Sawada and A. Sugamoto (Tsukuba 1979).

\bibitem{grs1979}
M. Gell-Mann, P. Ramond, and R. Slansky, {\it Supergravity}, ed. F. van Nieuwenhuizen and D. Freedman
(North Holland 1979).

\bibitem{ms1980}
R.N. Mohapatra and G. Senjanovi\'{c}, Phys. Rev. Lett. \textbf{44}, 912 (1980).





\bibitem{fy1986}
M. Fukugita and T. Yanagida, Phys. Lett. B \textbf{174}, 45 (1986).





\bibitem{lpy1986}
P. Langacker, R.D. Peccei, and T. Yanagida, Mod. Phys. Lett. A
\textbf{1}, 541 (1986); M.A. Luty, Phys. Rev. D \textbf{45}, 455
(1992); R.N. Mohapatra and X. Zhang, Phys. Rev. D \textbf{46}, 5331 (1992).





\bibitem{fps1995}
M. Flanz, E.A. Paschos, and U. Sarkar, Phys. Lett. B \textbf{345},
248 (1995); M. Flanz, E.A. Paschos, U. Sarkar, and J. Weiss, Phys.
Lett. B \textbf{389}, 693 (1996); L. Covi, E. Roulet, and F.
Vissani, Phys. Lett. B \textbf{384}, 169 (1996); A. Pilaftsis, Phys.
Rev. D \textbf{56}, 5431 (1997).


\bibitem{ms1998}
E. Ma and U. Sarkar, Phys. Rev. Lett. \textbf{80}, 5716 (1998).


\bibitem{bcst1999}
R. Barbieri, P. Creminelli, A. Strumia, and N. Tetradis, Nucl. Phys. B \textbf{575}, 61 (2000).




\bibitem{hambye2001}
T. Hambye, Nucl. Phys. B \textbf{633}, 171 (2002).


\bibitem{di2002}
S. Davidson and A. Ibarra, Phys. Lett. B \textbf{535}, 25 (2002); W.
Buchm\"{u}ller, P. Di Bari, and M. Pl\"{u}macher, Nucl. Phys. B
\textbf{665}, 445 (2003).

\bibitem{gnrrs2003}
G.F. Giudice, A. Notari, M. Raidal, A. Riotto, and A. Strumia, Nucl. Phys. B \textbf{685}, 89 (2004).


\bibitem{hs2004}
T. Hambye and G. Senjanovi\'{c}, Phys. Lett. B \textbf{582}, 73
(2004); S. Antusch and S.F. King, Phys. Lett. B \textbf{597}, 199
(2004).


\bibitem{bbp2005}
W. Buchmuller, P. Di Bari, and M. Plumacher, Annals Phys. \textbf{315}, 305 (2005).



\bibitem{di2001}
S. Davidson and A. Ibarra, Nucl. Phys. B \textbf{618}, 171 (2001).




\bibitem{barr2004}
S.M. Barr, Phys. Rev. Lett. \textbf{92}, 101601 (2004). 



\bibitem{gh2016}
P.H. Gu and X.G. He, Eur. Phys. J. C \textbf{76}, 515 (2016).





\bibitem{ps1974}
J.C. Pati and A. Salam, Phys. Rev. D \textbf{10}, 275 (1974). 

\bibitem{mp1975}
R.N. Mohapatra and J.C. Pati, Phys. Rev. D \textbf{11}, 566 (1975).


\bibitem{mp1975-2}
R.N. Mohapatra and J.C. Pati, Phys. Rev. D \textbf{11}, 2558 (1975). 


\bibitem{ms1975}
R.N. Mohapatra and G. Senjanovi\'{c}, Phys. Rev. D \textbf{12}, 1502 (1975).





\bibitem{krs1985}
V.A. Kuzmin, V.A. Rubakov, and M.E. Shaposhnikov, Phys. Lett. B \textbf{155}, 36 (1985).







\bibitem{bm1989}
K.S. Babu and R.N. Mohapatra, Phys. Rev. Lett. \textbf{62}, 1079 (1989).





\bibitem{bm1990}
K.S. Babu and R.N. Mohapatra, Phys. Rev. D \textbf{41}, 1286 (1990). 

\bibitem{bcs1991}
S.M. Barr, D. Chang, and G. Senjanovi\'{c}, Phys. Rev. Lett. \textbf{67}, 2765 (1991).



\bibitem{gu2012}
P.H. Gu, Phys. Lett. B \textbf{713}, 485 (2012).



\bibitem{cgnr2013}
S. Chakdar, K. Ghosh, S. Nandi, and S. Rai, Phys. Rev. D \textbf{88}, 095005 (2013).




\bibitem{gu2014}
P.H. Gu, Phys. Rev. D \textbf{96}, 075024 (2017).




\bibitem{gh2016}
P.H. Gu and X.G. He, Eur. Phys. J. C \textbf{76}, 515 (2016).





\bibitem{abbas2016}
G. Abbas, Mod. Phys. Lett. A \textbf{31}, 1650117 (2016).


\bibitem{abbas2017}
G. Abbas, Phys. Rev. D \textbf{95}, 015029 (2017).

\bibitem{abbas2017-2}
G. Abbas, Mod. Phys. Lett. A \textbf{34},1950119 (2019). 

\bibitem{abbas2017-3}
G. Abbas, Phys. Lett. B \textbf{773}, 252 (2017).



\bibitem{gu2017-2}
P.H. Gu, JHEP \textbf{1710}, 016 (2017).








\bibitem{gu2017}
P.H. Gu, Phys. Rev. D \textbf{96}, 055038 (2017).



\bibitem{fy1990}
M. Fukugita and T. Yanagida, Phys. Rev. D \textbf{42}, 1285 (1990).




 
  \bibitem{ht1990}
J.A. Harvey and M.S. Turner, Phys. Rev. D \textbf{42}, 3344 (1990).




\bibitem{kt1990}
E.W. Kolb and M.S. Turner, \textit{The Early Universe},
Addison-Wesley, 1990.




\end{thebibliography}
\end{document}